\documentclass[12pt]{article}

\usepackage{scicite}

\usepackage{times}

\usepackage{amsmath}
\usepackage{amsfonts}
\usepackage{amssymb}
\usepackage{graphicx}
\newcommand{\spp}{sp$^2$}
\usepackage[version=3]{mhchem}
\usepackage{gensymb}
\usepackage[labelfont=bf]{caption}

\usepackage[utf8]{inputenc}
\usepackage[english]{babel}
\usepackage[usenames, dvipsnames]{color}

\newenvironment{sciabstract}{%
\begin{quote} }
{\end{quote}}

\title{Quantitative principles for precise engineering of sensitivity in carbon-based electrochemical sensors} 

\author
{Ting Wu,${}^{1\dagger}$ Abdullah Alharbi,${}^{1\dagger}$ Roozbeh Kiani${}^{2,3}$, Davood Shahrjerdi${}^{1,4\ast}$\\
\\
\normalsize{${}^{1}$Electrical and Computer Engineering, New York University, Brooklyn, NY 11201}\\
\normalsize{${}^{2}$Center for Neural Science, New York University, New York, NY, 10003, USA}\\
\normalsize{${}^{3}$Neuroscience Institute, New York University Langone Medical Center, New York, NY 10016}\\
\normalsize{${}^{4}$Center for Quantum Phenomena, Physics Department, New York University, NY 10003}\\
\\
\normalsize{$^\ast$To whom correspondence should be addressed; E-mail:  davood@nyu.edu.}\\
\normalsize{$^\dagger$These authors contributed equally to this work.}
}

\date{}

\begin{document} 


\baselineskip18pt 


\maketitle 



\begin{sciabstract}
A major practical barrier for implementing carbon-based electrode arrays with high device-packing density is to ensure large, predictable, and homogeneous sensitivities across the array. Overcoming this barrier depends on quantitative models to predict electrode sensitivity from its material structure. However, such models are currently lacking. Here, we show that the sensitivity of multilayer graphene electrodes increases linearly with the average point defect density, whereas it is unaffected by line defects or oxygen-containing groups. These quantitative relationships persist until the electrode material transitions to a fully disordered \spp~carbon, where sensitivity declines sharply. We show that our results generalize to a variety of graphene production methods and use them to derive a predictive model that guides nano-engineering graphene structure for optimum sensitivity. Our approach achieves reproducible fabrication of miniaturized sensors with extraordinarily higher sensitivity than conventional material. These results lay the foundation for new integrated electrochemical sensor arrays based on nano-engineered graphene.
\end{sciabstract}
\clearpage
\baselineskip24pt

The ease of fabrication and operation of carbon-based electrochemical sensors gives them the potential to enable a new class of integrated sensor systems with wide-ranging applications from drug development to clinical diagnostics. To support these applications, the sensor system requires high spatial density (\textit{i.e.}, a dense packing of \textit{miniaturized} sensors) and consistent operations across the sensor array (\textit{i.e.}, sensors with predictable and homogeneous sensitivity). Moreover, high-precision applications additionally require electrodes with high sensitivity. Although the availability of advanced fabrication techniques would allow miniaturization of carbon-based electrochemical sensors~\cite{gao2016fully,xuan2016fully,nasri2017hybrid,xuan2018wearable}, satisfying the low variability and predictability requirements of sensitivity across a dense sensor array remains a difficult challenge. The sensitivity of this family of sensors is tied to the structural properties of the electrode material~\cite{shao2010graphene,jacobs2010carbon,zhu2017overview}. However, it is natural for the material to have atomic-level structural inhomogeneity, which can cause variability in the electrode sensitivity among sensors. Due to the random spatial distribution of the structural inhomogeneities in the electrode material, this variability becomes more pronounced with reducing the sensor size. Further, no method is currently available for predicting the electrode sensitivity from its material structure. The current limitations in producing predictable and homogeneous array of electrodes with high sensitivity have prevented the implementation of highly integrated sensor systems based on carbon-based electrochemical sensors.

To account for the variability and also determine the electrode sensitivity, the common practice is to calibrate each sensor through post-manufacturing measurements, which involves creating ``calibration curves" by measuring the sensor response to known concentrations of analytes~\cite{sinkala2012electrode}. Although this strategy is applicable for dealing with individual devices or a small sensor array, it is highly inefficient for the implementation of large-scale integrated sensor systems. A more tractable approach is to precisely engineer the material synthesis in order to reliably yield miniaturized electrodes with predictable and homogeneous sensitivities. The efficacy of this approach, however, hinges on an understanding of the quantitative relationship between electrode sensitivity and the structural properties of the material. Such a quantitative model is currently lacking. Our study aims to provide this model.

Among different forms of carbon, electrodes from \spp~carbon nanomaterials (\textit{e.g.}, carbon nanotubes, graphene, and reduced graphene oxide) have shown high sensitivity~\cite{schmidt2013carbon,keeley2014electroanalytical,gao2015effect,chung2013biomedical}, making them promising for sensing applications that might require high precision measurements of an analyte concentration. The sensitivity of those electrodes has been often attributed to structural defects such as edge plane sites~\cite{banks2006new,banks2005electrocatalysis,schmidt2013carbon} or oxygen-containing functional groups~\cite{gao2015effect,keeley2014electroanalytical}. However\textemdash and this attests to the challenging nature of the problem\textemdash no study has yet directly tested and quantified the effect of different defect types on the sensitivity of carbon-based electrodes. Overcoming this problem will result in a quantitative understanding of how structural defects influence electrode sensitivity\textemdash a crucial building block for a desired predictive model. 

To address this problem, we undertake a systematic study of sensitivity in miniaturized carbon-based electrochemical sensors consisting of multilayer graphene electrodes. We used multilayer graphene as a model system due to the ease of defect engineering~\cite{ugeda2010missing} and the availability of experimentally-established Raman-based models for quantifying defects~\cite{tuinstra1970raman,canccado2006general,lucchese2010quantifying,canccado2011quantifying,canccado2017disentangling}. Our tight control over the sensor fabrication process, together with the accuracy and reproducibility of our material characterizations and sensor measurements, have allowed us, for the first time, to experimentally determine the quantitative relationships between various defect types and the electrode sensitivity in graphene-based electrochemical sensors. 

We find that the electrode sensitivity is amplified in linear proportion to the average density of point defects. In contrast, the average density of line defects or oxygen-containing groups has no effect on the electrode sensitivity. Once the concentration of point defects exceeds a threshold value, which indicates the onset of transition to a fully disordered \spp~carbon material, electrode sensitivity decreases rapidly. Based on these findings, we derive a predictive model for engineering the structural properties of carbon-based electrodes on an atomic level to match a desired sensitivity. As a practical example for the utility of our model and methods, we nano-engineer electrodes with unprecedented sensitivity for \textit{in vitro} measurements of dopamine. Our engineered electrodes show up to 20 times higher sensitivity to dopamine than conventional carbon fiber (CF) electrodes~\cite{wightman1988detection,wightman2002transient,kishida2016subsecond}. Moreover, we show that the sensitivity of our engineered electrodes match the predictions of our quantitative model. Finally, we show that our model consistently applies to graphene-based sensors produced through different synthesis methods, promising wide applicability of our findings for prediction and engineering of the sensitivity of carbon-based electrochemical sensors. 

\subsection*{Results}
\paragraph*{Enhanced electrode sensitivity in multilayer graphene.} To evaluate the link between the atomic structure of multilayer graphene and the sensitivity of electrodes made of it, we performed electrochemical sensing of biochemical molecules using fast-scan cyclic voltammetry (FSCV). Owing to its good ionic specificity and sub-second detection ability, FSCV with carbon-based electrodes has been used extensively for measuring biochemical molecules in chemically complex environments such as the brain~\cite{kuhr1986real,wightman1988detection,BJVenton,schmidt2013carbon}. We initially constructed FSCV electrodes using multilayer graphene grown by chemical vapor deposition (CVD) and measured their sensitivity for neurochemical molecules \textit{in vitro}. CVD graphene films typically have different amounts of \spp-hybridized defects, due to minor differences in the production method, apparatus, or even the granular structure of the growth substrate~\cite{reina2008large,li2011large}. To increase the diversity of different defect densities in our sensor electrodes, we obtained several batches of CVD multilayer graphene films grown on nickle foils. 

\begin{figure}[t!]
  \centering
  \includegraphics{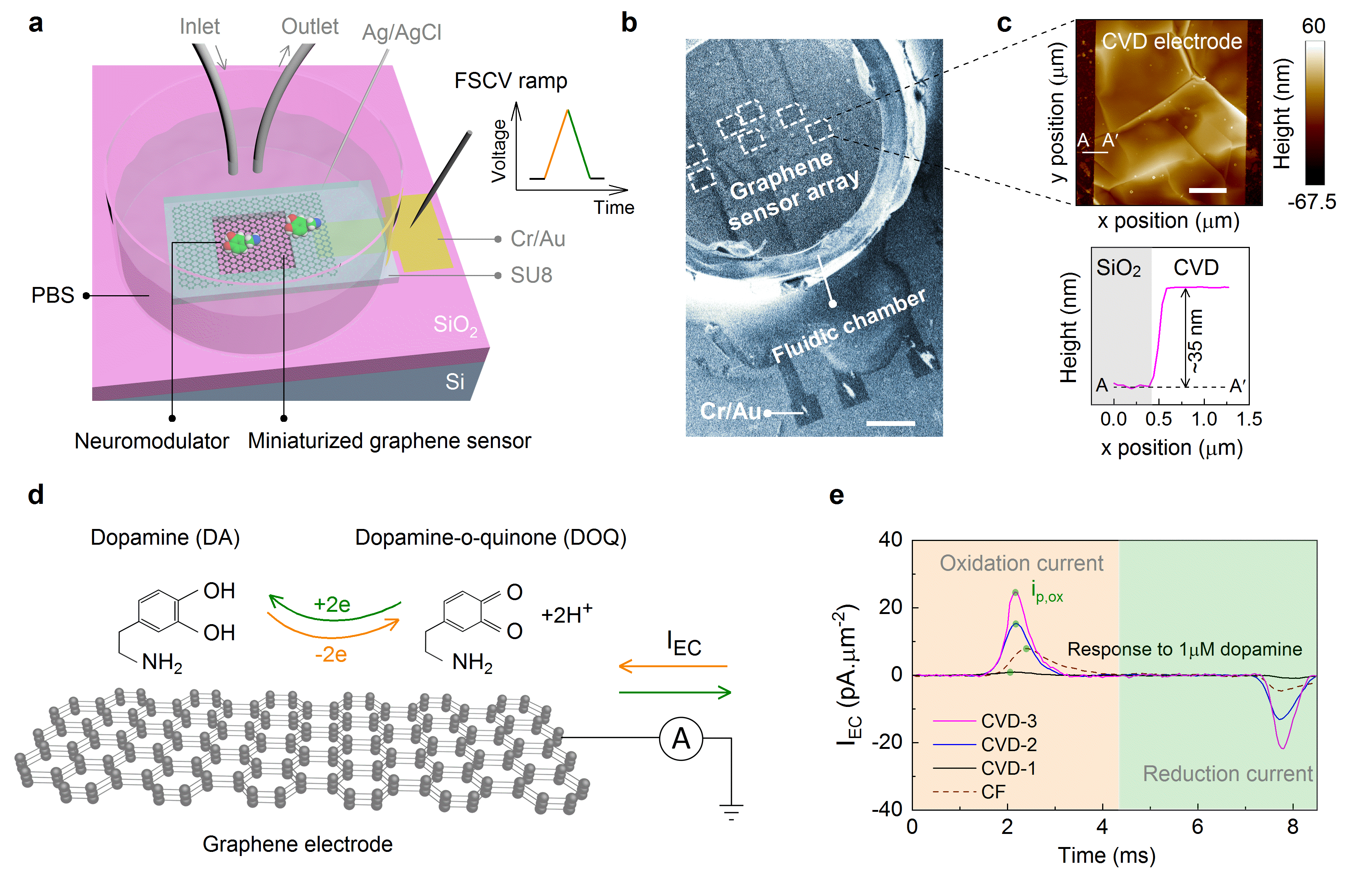}
  \caption{\textbf{FSCV sensors made of multilayer graphene films with different amounts of structural defects.} (a) Schematics of a graphene-based electrode used for FSCV measurements. The multilayer graphene electrode was mounted on a \ce{SiO2}/Si substrate and connected to a Cr/Au contact. A fluidic chamber filled with PBS solution of target biochemicals (dopamine or serotonin neuromodulators) was made around the sensor. To maintain neuromodulator concentration at a desired level, fresh solution was brought to the chamber by an inlet and old solution was taken out by an outlet. (b) The SEM image shows an example of our miniaturized graphene-based sensor array and fluidic chamber around the sensors. We used nanofabrication to build miniaturized sensors from our candidate CVD and graphitized films. Scale bar is 300 $\mu$m. (c) Topographic image of an example CVD multillayer graphene sensor and its thickness measured by atomic force microscopy. Scale bar is 5 $\mu$m. (d) In FSCV measurements of dopamine, the voltage is applied to the sensor electrode; it first quickly ramps up, which oxidizes dopamine to dopamine-o-quinone, and then ramps down, which reduces it back to dopamine. The resulting current is measured. (e) Area-normalized electrochemical current (I$_{EC}$) as a function of time in one FSCV cycle for four sample electrodes made of CVD multilayer graphene films and carbon fibers. The noticeable variations of I$_{EC}$ for these sensors in response to the same dopamine concentration highlight the critical role of structural defects on sensitivity. \label{fig1}}
\end{figure}

To fabricate electrodes, we transferred the CVD multilayer graphene films with an average thickness of 30 nm onto \ce{SiO2}/Si substrates using standard chemical layer-transfer processes~\cite{li2009transfer,kim2009large,lee2010wafer}. Using nanofabrication, we then made miniaturized sensor electrodes with a planar geometry, shown schematically in Fig.~\ref{fig1}a. The details of the fabrication process are given in Supporting Information. We used a similar process for fabricating all devices discussed in this paper. We designed the fabrication process around two key factors. The first one is to avoid creating unintentional defects in multilayer graphene during the sensor fabrication. This consideration is particularly important when making electrodes from defect-engineered multilayer graphene, discussed later. To do so, during the fabrication process, we protected the sensing region of the electrode with a thin metal layer (Cr/Au; 5 nm/50 nm). Second, for analyzing the sensor response in our study, we used the area-normalized sensitivity. We defined the sensing region of the electrodes accurately using an SU8 encapsulating layer. This layer also protected the metal contact from exposure to the electrolyte solution. To perform the sensing experiments, we removed the protective metal stack and mounted a fluidic chamber on the samples. Fig.~\ref{fig1}b shows the top-view scanning electron microscopy (SEM) image of an example sensor array. For comparison, we also fabricated electrodes from conventional CFs (Fig. S3), commonly used in FSCV measurements of neurochemicals in the brain~\cite{wightman2002transient,bucher2015electrochemical}.

We characterized the sensitivity of the fabricated sensors through FSCV measurements of dopamine\textemdash an important neuromodulator for action-selection and reward-motivated behavior~\cite{Glimcher13092011,SB-Floresco,RM-Costa}. During the FSCV measurement, dopamine undergoes a redox reaction (Fig.~\ref{fig1}d), where it is oxidized to dopamine-o-quinone (DOQ) by a voltage ramp-up applied to the electrode (see Fig.~\ref{fig1}a). The amplitude of the resulting oxidation current is a measure of the dopamine concentration. The voltage subsequently ramps down, causing the DOQ molecules to be reduced back to dopamine, which gives rise to a reduction current. FSCV estimates dopamine concentration based on the magnitude of the oxidation current. 
Electrode sensitivity represents the change in the peak of the oxidation current (i$_{p,ox}$) per unit concentration of a biomolecule (Fig. S13c-d). We defined unit concentration as $\mu$M and the area-normalized sensitivity, S$_A$, as i$_{p,ox}$ at 1 $\mu$M divided by electrode area. Because the amplitude of the electrochemical current is proportional to the geometric surface area of the sensors, normalization of sensitivity by sensor area enables comparison of sensors with diverse sizes. Surface roughness increases the geometric surface area and can potentially bias the area-normalized sensitivity. Therefore, we estimated the total surface area of our multilayer graphene sensors by performing atomic force microscopy (AFM) and measuring surface roughness before FSCV experiments (Fig.~\ref{fig1}c, Fig. S16-S17). As a result, our area-normalized sensitivity is independent of the sensor geometry and reflects the inherent sensing property of the electrode material. 

Fig.~\ref{fig1}e shows the area-normalized electrochemical current (I$_{EC}$) curves for four example electrodes (three CVD and one CF) in response to a 1 $\mu$M dopamine solution. The green circles on the curves denote i$_{p,ox}$. This plot and the FSCV measurements of the other CVD electrodes (Fig. S14) demonstrate substantial variations in electrode sensitivity. Many sensors were minimally responsive to dopamine molecules and a few showed noticeably higher S$_A$ than the CF devices. We hypothesized that diversity of structural defects of the sensing material was crucial for explaining the wide range of observed electrode sensitivities. 

\begin{figure}[t!]
  \centering
  \includegraphics{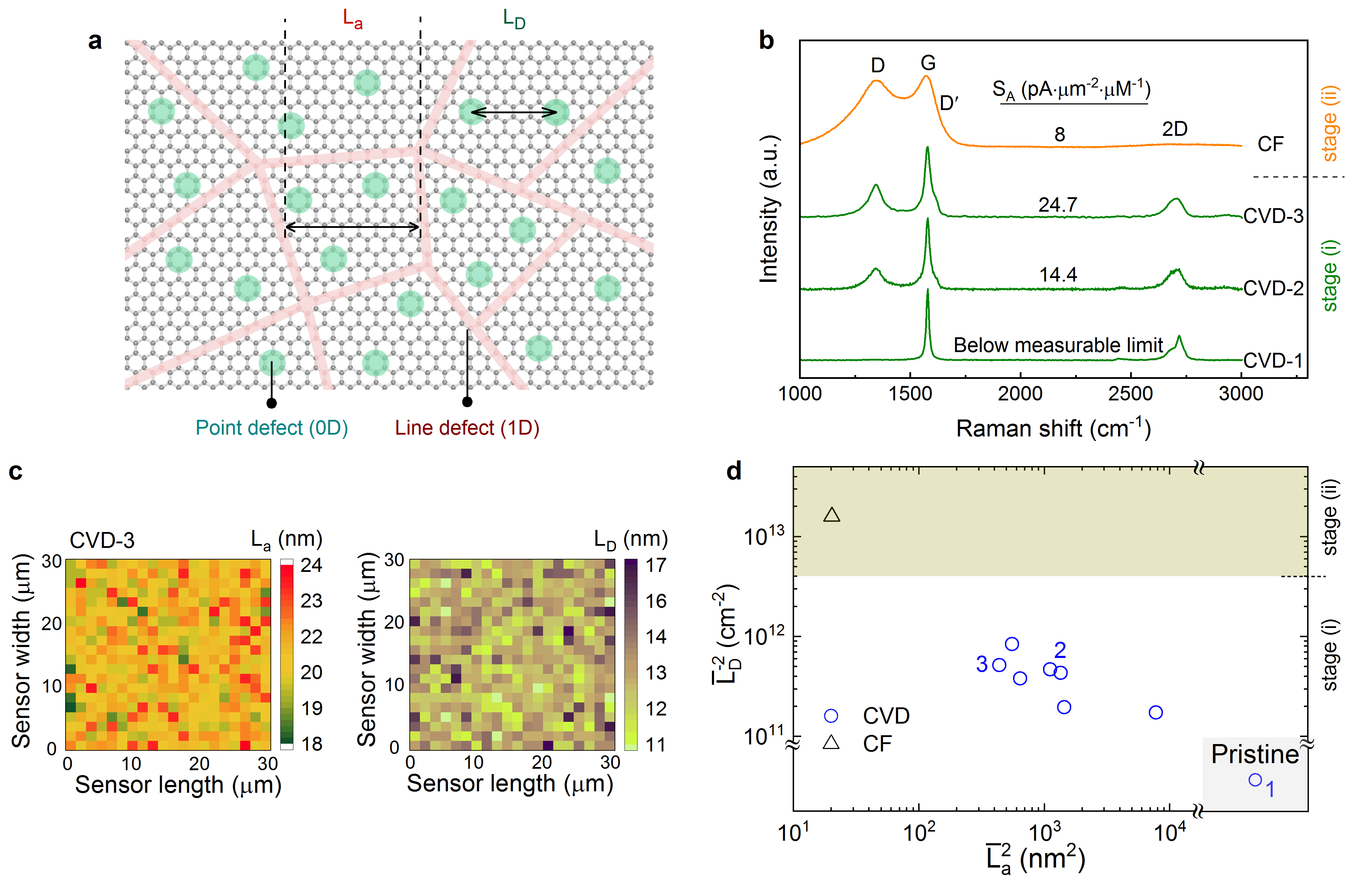}
  \caption{\textbf{Quantifying structural defects in graphene electrodes.} (a) Line and point defects are typically present at the same time in synthetic multilayer graphene films and can be characterized by the average crystallite size, L$_a$, and average distance between point defects, L$_D$, respectively. (b) To evaluate the structural properties of the CVD films, we used Raman spectroscopy. The increase of the D peak intensity for samples from bottom to top indicates the higher density of \spp-hybridized defects. The gradual changes of the Raman peaks also highlight the transition from a graphitic structure in stage (i) into a fully disordered \spp~carbon in stage (ii). Conventional CFs are typically in stage (ii) as shown by the example electrode in the top row. Area-normalized sensitivity, S$_A$, is indicated for each sample electrode. (c) Spatially resolved L$_a$ and L$_D$ across the sensor surface for an example CVD electrode. To estimate the average density of point and line defects, we obtained similar spatial maps for all electrodes studied in this work. (d) The scatter plot of the average crystallite area ($\overline{L}_a^{2}$) and the average point defect density ($\overline{L}_D^{-2}$) shows that our candidate materials covered a broad range of defect densities. Numbers next to the CVD samples indicate example electrodes in panel b. The yellow shading represents the stage (ii) of the amorphization trajectory, while the gray box marks the detection limit of Raman for estimating the point and line defect concentration.  \label{fig2}}
\end{figure}

\paragraph*{Quantifying structural defects in carbon-based electrodes.} The ability to distinguish different types of defects and quantify their amounts in the electrode material not only is critical for revealing the connection between structural defects and the electrode sensitivity, but might also lead to a predicative framework for the sensitivity-targeted engineering of the sensor material. While the types of structural defects are diverse, one simple way to classify them is based on their dimensionality. For example, defects in materials with a two-dimensional lattice, such as graphene, are either zero-dimensional (point defects) or one-dimensional (line defects). Examples of point defects in graphene are vacancies~\cite{lee2005diffusion,vicarelli2015controlling,banhart2010structural}, Stone-Wales defects~\cite{banhart2010structural,vicarelli2015controlling}, and dopants~\cite{zhang2012nitrogen}. On the other hand, crystallite borders~\cite{canccado2006general} and extended dislocations~\cite{vicarelli2015controlling} are examples of line defects. Point and line defects are often simultaneously present in synthetic graphene-based materials, as shown schematically in Fig.~\ref{fig2}a. Raman spectroscopy has been experimentally established as a useful non-destructive method for identifying and quantifying \spp-hybridized defects in graphene based on their dimensionality~\cite{tuinstra1970raman,canccado2006general,lucchese2010quantifying,canccado2011quantifying,canccado2017disentangling}. 

Fig.~\ref{fig2}b shows representative Raman spectra for a few CVD samples and a CF electrode. The distinct peaks in the Raman spectrum of multilayer graphene films are well-studied~\cite{ferrari2007raman,ferrari2013raman}. The G peak appears at about 1576 cm$^{-1}$ and signifies the \spp~hybridization of carbon atoms. The D peak arises from the breathing modes of aromatic carbon rings and signifies \spp-hybridized defects. The 2D peak is the second-order of the D peak, which is present only in fully \spp-bonded carbon materials. Changes of these peaks in Fig.~\ref{fig2}b (from bottom to top) illustrates the gradual transition of the film structure (i) from a highly ordered multilayer graphene to a disordered nanocrystalline graphite and (ii) finally to a fully disordered \spp~carbon material. In stage (i), the D peak intensity increases monotonically and the 2D peak is visible in the Raman spectra. Upon transition into stage (ii), the 2D peak becomes noticeably broad and its intensity weakens dramatically. The combination of our CVD and CF electrodes covered the whole spectrum of the graphene amorphization trajectory. 

We applied a recently advanced theoretical method~\cite{canccado2017disentangling} for extracting the amounts of point and line defects from the measured Raman spectra of our sensor samples (see Supporting Information). The results of this method have been previously validated by scanning tunneling microscopy~\cite{canccado2017disentangling}, illustrating its ability to unambiguously distinguish point and line defects in graphene-based samples. This theoretical method relies on numerical simulations based on the area ratio of the D and G peaks and the line width of the G-band to derive the average crystallite size (L$_a$) and the average distance between point defects (L$_D$) within the spot size of the Raman laser. The details of our L$_a$ and L$_D$ calculations are given in Supporting Information. We note that this methodology for quantifying defects is independent of the production method of multilayer graphene, making it suitable for our study involving CVD, graphitized, and CF materials.    

Since the location of defects on a sensor electrode is random, we estimated the concentration of each defect type on a sensor electrode by measuring the number of defects averaged over the sensor surface area. To do so, we first obtained the spatial Raman maps of our sensor electrodes and quantified L$_a$ and L$_D$ at each Raman spot (see Supporting Information). Fig.~\ref{fig2}c shows the spatially resolved distributions of L$_a$ and L$_D$ for an example CVD electrode. The mean values from the L$_a$ and L$_D$ distributions were then used for estimating the average density of point defects ($\overline{L_D}^{-2}$) and average crystallite area ($\overline{L_a}^2$) in our sensor samples. This methodology allows us to analyze the relationship between the area-normalized sensitivity and the concentration of each defect type in our carbon-based electrodes.  

Fig.~\ref{fig2}d shows the scatter of $\overline{L_D}^{-2}$ and $\overline{L_a}^2$ for our CVD sensor samples, highlighting the large diversity of line and point defects in our candidate sensor samples. In this plot, the gray box (in the lower right corner) marks the region, where Raman lacks accuracy for estimating point and line defect concentrations, because the expected L$_D$ and L$_a$ values are beyond the upper detection limits of Raman. The Raman spectra of samples that fall in this region typically do not show a visible D peak. We refer to those samples as pristine. Moreover, past Raman studies of defective graphene suggest that the onset of stage (ii) of the amorphization trajectory occurs at L$_D$ of about 4-5 nm~\cite{canccado2011quantifying}. This length scale is comparable to the localization length of the disorder-induced Raman D band at 300 K~\cite{beams2011low}. The yellow shading in Fig.~\ref{fig2}d marks the stage (ii), where the CF sample is. In contrast, our CVD sensor samples were in stage (i) of the amorphization trajectory. 

\begin{figure}[t!]
  \centering
  \includegraphics{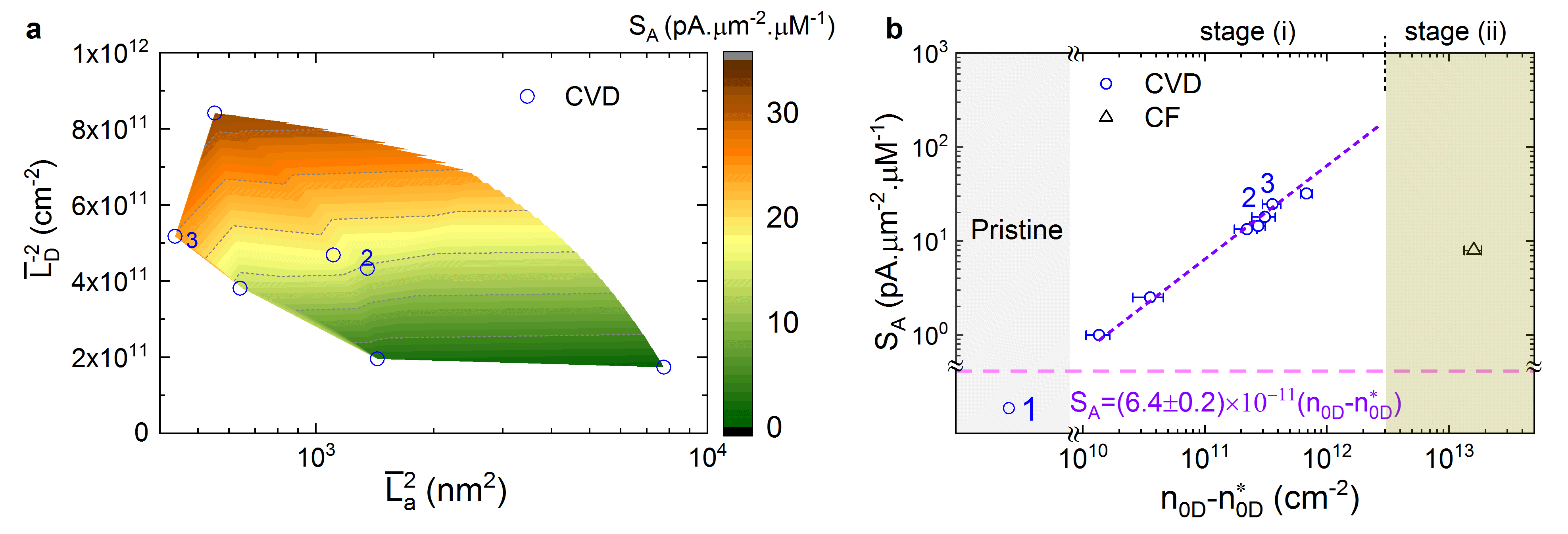}
  \caption{\textbf{Revealing the link between structural defects and electrode sensitivity.} (a) Contour plot of S$_A$ \textit{versus} $\overline{L}_a^{2}$ and $\overline{L}_D^{-2}$, indicating that the sensitivity of the CVD sensors in stage (i) was largely independent of the average density of line defects and was amplified by increasing the average density of point defects. (b) We found that S$_A$ of the CVD sensors in stage (i) was amplified in linear proportion to the density of point defects, and dropped upon transition into stage (ii) (yellow shading). The dashed line represents the measurable limit of S$_A$ in our sensor readout system. The sensitivity of electrodes from pristine graphene was below the measurable limit.} \label{fig3}
  
\end{figure}

\paragraph*{Effect of defects on electrode sensitivity.} To reveal the quantitative effect of defects on multilayer graphene sensors in stage (i) of the amorphization trajectory, we made a contour plot of the area-normalized sensitivity (S$_A$) as a function of the average crystallite area ($\overline{L_a}^2$) and the average point defect density ($\overline{L_D}^{-2}$), shown in Fig.~\ref{fig3}a. The plot shows that electrodes with similar density of point defects exhibited nearly similar S$_A$, regardless of their crystallite size. Further, S$_A$ was amplified with increasing the density of point defects. 

Next, we analyzed the relationship between S$_A$ and the point defect concentration (\textit{i.e.}, n$_{0D}$ = L$_D^{-2}$). Plotting S$_A$ as a function of the point defect concentration indicated a linear relation between the area-normalized sensitivity and the point defect density when the sensing material was in stage (i), as shown in Fig.~\ref{fig3}b. From the linear fit to the data, we found that the x-intercept occurs at a point defect density of 1.6 $\times$ 10$^{11}$ cm$^{-2}$. We denoted this point defect concentration as n$_{0D}^*$. From these observations, we then derived an empirical model that represents the experimentally measured S$_A$ for dopamine when the electrode material was in stage (i) of the amorphization trajectory. This model, given below, is valid for point defect concentrations greater than n$_{0D}^*$:
\begin{align}
 S_A = (6.4 \pm 0.2) \times 10^{-11} (n_{0D} - n_{0D}^*)
  \label{model}
\end{align}
where S$_A$ and n$_{0D}$ have units of pA.$\mu$m$^{-2}$.$\mu$M$^{-1}$ and cm$^{-2}$, respectively. We note that in our sensing experiments, electrodes containing a point defect concentration below n$_{0D}^*$, including those made from pristine graphene, are minimally responsive, \textit{i.e.}, their sensitivity was below the measurable limit of our readout system (the magenta dashed line in Fig.~\ref{fig3}b). From the data, we also found that upon transition into stage (ii), where the CF electrode is, S$_A$ no longer follows the linear trend. 

On the basis of the above observations, we made the following hypotheses regarding the sensitivity-targeted engineering of the graphene synthesis process. First, the synthesis process does not need to be optimized for line defects, since they had no observable effect on the electrode sensitivity of CVD sensor samples. Second, the quantitative relationship between S$_A$ and the density of point defects in stage (i), quantitatively described by equation \eqref{model}, can be used as a predictive model for tuning the sensitivity of electrodes. And finally, the sensitivity degrades as the material structure transitions into stage (ii). An important prediction of these hypotheses is that S$_A$ can be maximized by maximizing the density of point defects in multilayer graphene before transitioning to a fully disordered \spp~carbon material. 

\begin{figure}[t!]
  \centering
   \includegraphics[width=\textwidth]{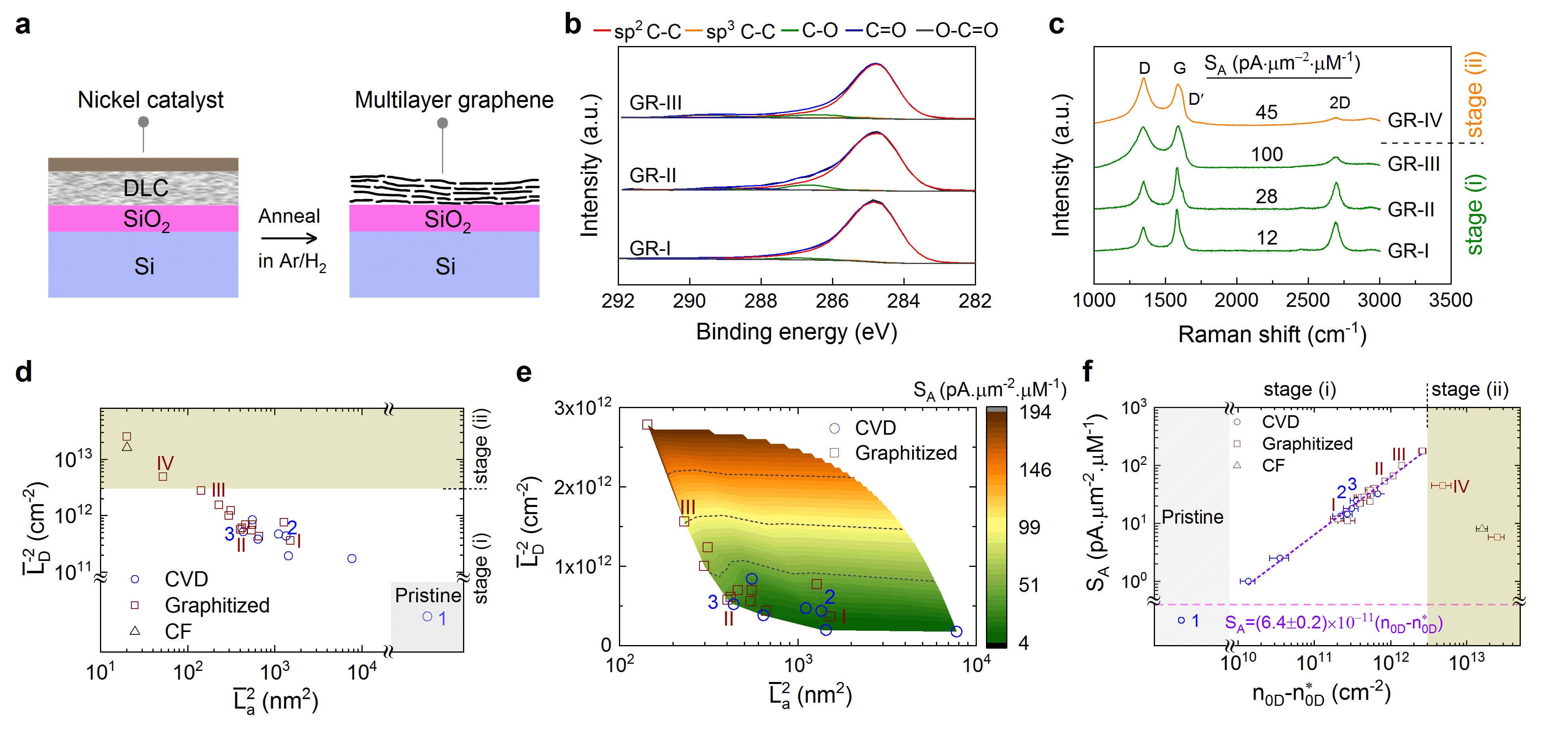}
  \caption{\textbf{Engineered multilayer graphene sensors with predictable sensitivity.} (a) To confirm the connection between the area-normalized sensitivity and the density of point defects, we produced engineered multilayer graphene through nickel-induced graphitization of diamond-like carbon (DLC). (b) XPS measurements confirmed the graphitic nature of the films through the presence of \spp-hybridized C-C bond peak highlighted in red. (c) We used Raman measurements to quantify the structural defects in our graphitized samples. (d) We created three groups of graphitized samples. A first group that had overlapping density of point defects with the CVD sensors; a second group which contained a higher density of point defects than the CVD samples without transitioning to stage (ii); and a third group that had even higher density of point defects and were in stage (ii) (yellow shading). Numbers next to the data points denote the example GR (I,II,III,IV) and CVD (1,2,3) electrodes in panel c and Fig.~\ref{fig1}b, respectively. (e) From the FSCV measurements of dopamine, we confirmed that the electrode sensitivity of the GR samples was largely independent of the crystallite size and increased with the point defect concentrations. (f) We found that GR sensor samples with a similar density of point defects to CVD samples had the same S$_A$. By increasing the density of point defects in stage (i), we achieved a maximum S$_A$ of about 177 pA.$\mu$m$^{-2}$.$\mu$M$^{-1}$, which is 20 times higher than conventional CFs. The electrode sensitivity decreased rapidly once the structure of the carbon lattice transitioned into stage (ii). The dashed line denotes the minimum measurable limit of S$_A$ in our experiments. \label{fig4}}
\end{figure}

\paragraph*{Engineered electrodes with predictable sensitivity.} We directly tested the validity of our hypotheses by developing a process that allowed us to engineer multilayer graphene films with different amounts of point and line defects. Of the various approaches for producing multilayer graphene~\cite{reina2008large,Rodriguez15012013,wang2011ni,hass2008growth, kim2009large, cao2010large}, we adapted a method based on metal-induced transformation of amorphous carbon to multilayer graphene using a thin nickel catalyst~\cite{rodriguez2011graphene,berman2016metal}. Our process involved creating diamond-like carbon (DLC) islands directly on \ce{SiO2}/Si substrates and graphitizing them at temperatures between 1000 and 1100 \degree C, as shown in Fig.~\ref{fig4}a (see Supporting Information for details of the graphitization process). We probed the graphitic structure of the samples using high-resolution x-ray photoelectron spectroscopy (XPS) (Fig.~\ref{fig4}b). Curve fitting of the XPS data indicated the \spp~nature of the carbon-carbon bonds. The 2D peaks in Raman spectra of the films (Fig.~\ref{fig4}c) further confirmed the multilayer graphene growth. We observed that the structural properties of the multilayer graphene films made from this method sensitively depended on the nickel thickness, annealing temperature, and growth time. By tuning these parameters, guided by Raman analysis, we could reliably and reproducibly generate samples with desired amounts of structural defects. We refer to these multilayer graphene samples as the \textit{graphitized} (GR) samples. 

In our experiments, we created three sets of graphitized sensor samples. A first group that contained the same amounts of point defects as the CVD samples. This was to reproduce the sensitivity of those CVD electrodes and hence to demonstrate that our experimental predictive model of electrode sensitivity is independent of the graphene production method. A second group of samples had significantly higher amounts of point defects than the CVD devices without transitioning to stage (ii). This set of GR samples not only allowed us to investigate the validity of our predictive model for the range of point defect concentrations that was not covered by the CVD sensor samples, but also enabled us to explore the upper limit of S$_A$ in stage (i). We also created a third group of GR samples, which were in stage (ii) of the amorphization trajectory of graphene. These samples allowed us to confirm the drop in S$_A$ once the structure of the electrode material transforms into a fully disordered \spp~carbon material. Fig.~\ref{fig4}c shows the representative Raman spectra of a few GR samples and their corresponding S$_A$ for dopamine. In Fig.~\ref{fig4}d, we show the summary of average point and line defect concentrations for our engineered GR samples. In this plot, we have also included the CVD and CF samples to facilitate the distinction between the three groups of the GR samples. In addition to regions of overlap with the CVD and CF samples, note the region where the GR samples were engineered to have a significantly higher point defect density than the CVD samples without transitioning to stage (ii), the yellow shading in the plot.

We used miniaturized electrodes fabricated from defect-engineered graphitized films to perform FSCV measurements of dopamine (Fig. S15). Fig.~\ref{fig4}e shows the contour plot of S$_A$ for the graphitized and CVD electrodes, where the multilayer graphene sensor electrodes are in stage (i). The plot shows that like the CVD samples, the sensitivity of graphitized electrodes increased with point defect density and was independent of the crystallite size. Plotting the S$_A$ as a function of point defect density (see Fig. \ref{fig4}f) revealed three critical results. First, the S$_A$ of the graphitized sensors followed the same linear trend as the CVD samples, confirming that the average density of point defects was the main predictor of the S$_A$ for multilayer graphene films in stage (i). The exact method of production of the multilayer graphene film did not matter as long as the point defect concentration remained the same, hence confirming that our predictive model in equation \eqref{model} can be generalized to other graphene production methods. Second, by increasing the density of point defects in stage (i), our model predicts that one could maximize S$_A$. Indeed, we achieved a remarkably high S$_A$ of 177 pA.$\mu$m$^{-2}$.$\mu$M$^{-1}$ in response to dopamine, which is about 20 times higher than the sensitivities reported for CF electrodes in past studies~\cite{schluter2014real,zachek2009simultaneous}. Third, the transition of the multilayer graphene into a fully disordered \spp~carbon material caused a rapid degradation of S$_A$. Our data indicates that S$_A$ began to degrade at L$_D$ of about 5-6 nm, which coincides with the onset of transition to stage (ii). Conventional CF electrodes that are used for studying neurochemical changes in the brain~\cite{Robinson1763,wightman1988detection,schluter2014real,kishida2016subsecond} are usually in this regime. Our finding explains the fundamentally small sensitivity of these electrodes and suggests a practical method to significantly boost the electrode sensitivity for such measurements. 

To ensure our results were not limited to a particular molecule, we also performed \textit{in vitro} FSCV measurement of serotonin neurotransmitters using the multilayer graphene electrodes in stage (i). The results confirmed that the linear increase of electrode sensitivity with increasing average point defect density generalized to serotonin and was, therefore, a property of the electrode not the measured analyte (Fig. S18-S19).

\begin{figure}[t!]
  \centering
  \includegraphics{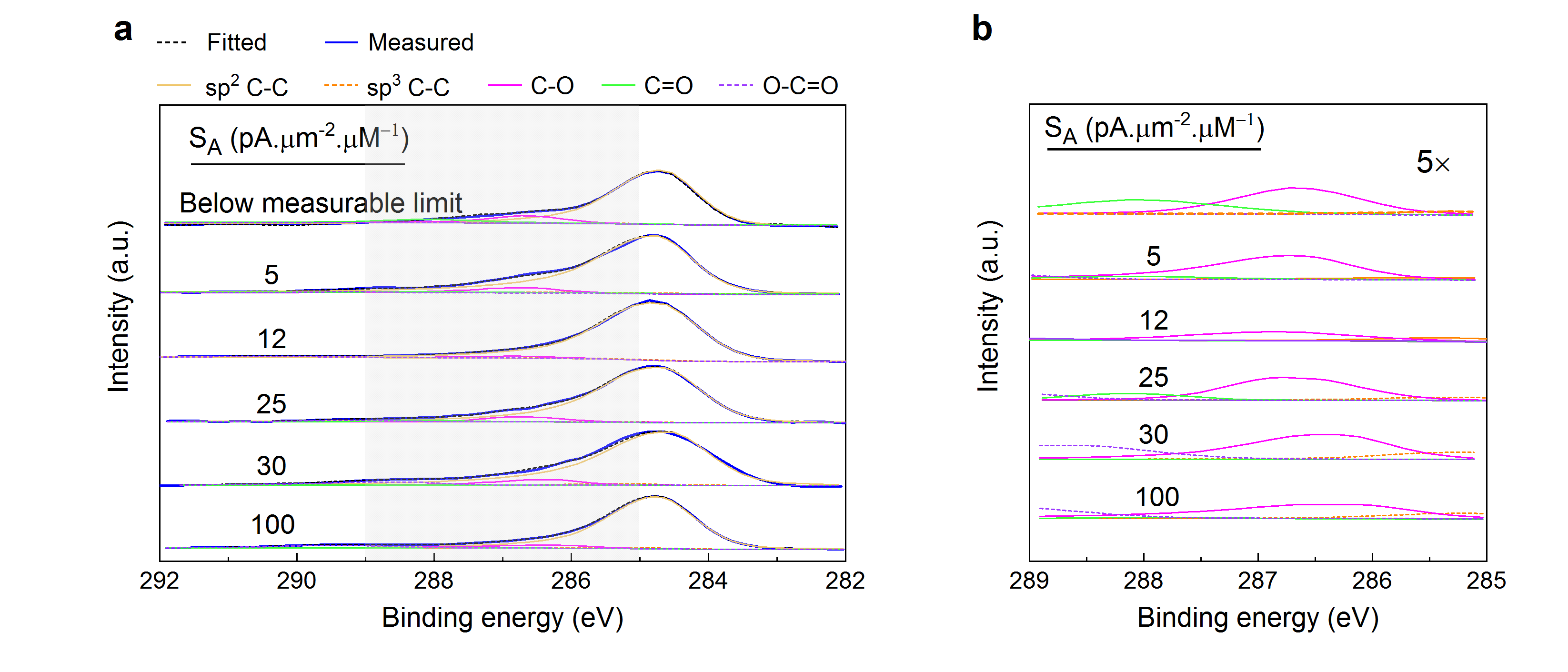}
  \caption{\textbf{Role of oxygen-containing functional groups on electrode sensitivity.} To investigate this, we performed XPS on multiple multilayer graphene electrodes with markedly different area-normalized sensitivity. (a) The XPS measurements were performed immediately after FSCV neurochemical measurements in a UHV environment. (b) The zoomed-in plot of the region highlighted by the gray shading in panel (a), showing the fitted curves for the different carbon-oxygen peaks. For better illustration, the intensity was also increased by a factor of 5. Side-by-side comparison of these samples revealed that oxygen-containing functional groups have no direct effect on the electrode sensitivity of multilayer graphene electrodes in stage (i).} \label{fig5}
  
\end{figure}

\paragraph*{Effect of oxygen-containing functional groups.} Previous studies hypothesized a direct correlation between the electrode sensitivity and the amount of oxygen-containing functional groups available on the surface of carbon electrodes~\cite{brajter2000nanostructured,schmidt2013carbon,yang2017o2}. To examine the role of oxygen functional groups, we performed XPS measurements on multiple multilayer graphene sensor electrodes with markedly different area-normalized sensitivity. The electrodes were chosen from both CVD and graphitized sensor samples and the XPS measurements were performed immediately after the FSCV experiments in an ultra high vacuum (UHV) environment. By analyzing the XPS data, shown in Fig.~\ref{fig5}, we ruled out a significant role of oxygen-containing functional groups in amplifying the sensitivity of our electrodes made from multilayer graphene. Hence, we conclude that point defects are the most critical factor for amplifying the sensitivity of graphene-based electrodes that are in stage (i) of the amorphization trajectory.

\subsection*{Discussion}

Our findings establish fundamental principles for predicting the sensitivity of graphene-based electrochemical sensors based on structural defects. Using these principles, we devised a quantitative methodology for reproducible fabrication of homogeneous sensors with optimized sensitivity. The remarkably high sensitivity of our miniaturized electrodes are due to the maximization of point defect concentration while keeping the electrode material in stage (i) of the amorphization trajectory of graphene.

The density of line defects and oxygen-containing functional groups appear to have minimal bearing on the electrode sensitivity. This observation simplifies the electrode manufacturing process by removing the need for monitoring and optimization of line defects and oxygen-containing groups. Further, it contradicts previous speculations that non-carbon functional groups may be key to the sensitivity of carbon-based electrodes~\cite{keeley2014electroanalytical,gao2015effect}. However, we do not rule out the possibility that one may further increase the sensitivity of our nanoengineered electrodes by adding special functional groups that amplify the effect of point defects.

It is currently unclear why point defects have such a critical role in determining the sensitivity of multilayer graphene electrodes. We speculate that these defects provide an optimal environment for electron transfer during the redox reaction by (i) fixing the chemical potential (Fermi energy) at the Dirac point~\cite{moktadir2015defect,massabeau2017evidence}, and (ii) simultaneously increasing the local density of states at about the Dirac point in proportion to their concentration in stage (i)~\cite{pereira2006disorder,ugeda2010missing}. However, the physical principles that govern this effect should yet be elucidated. Further, studying the correlation between the specific sub-types of point defects and the electrode sensitivity might be useful in further refining the theory of the sensor operation.     

Our discovery has far-reaching practical implications. By providing guidelines for optimizing the sensitivity of carbon-based electrochemical sensors, we enable significant improvements in a wide range of applications from a next generation of neural probes to multiplexed lab-on-a-chip sensing platforms, or more generally, wherever accurate recording of biochemical compounds using a dense array of sensors is essential. Specifically, our quantitative nanoengineering methodology can ensure fabrication of a dense array of sensors with predictable and homogeneous sensitivity, thus making such nanoengineered carbon-based electrodes suitable for compact, multi-channel sensor systems required in large-scale applications. Further, our nanoengineered electrodes overcome an existing obstacle for industrial-scale fabrication of reliable sensors with reproducible electrode sensitivity. Current methods for fabrication of carbon-based electrodes are not developed to optimize the density of point defects. Consequently, they are largely dependent on post-manufacturing measurement for calibration of sensors and are prone to producing minimally responsive sensors that have to be discarded. In contrast, our fabrication method enables industrial-scale and targeted nanomanufacturing of carbon-based electrodes that have desired sensitivity levels far beyond their predecessors. 

\section*{Materials and Methods}
\textbf{Multilayer graphene films:} CVD multilayer graphene samples were obtained from two commercial vendors: Graphene Supermarket and Graphene Platform, Inc. The CVD films were grown on nickel foils, which were then removed chemically during the layer transfer process using the commercial nickel etchant TFG (Transene Company Inc.). The free-standing CVD films were subsequently mounted on p$^+$ silicon substrates capped with 285 nm thermally grown \ce{SiO2}. The graphitized samples were produced using a custom-made system. Details of the metal-induced graphitization process is given in Supporting Information. 

\noindent \textbf{Raman measurements:} To quantify the structural defects in multilayer graphene films, Raman measurements were performed using Horiba Xplora $\mu$-Raman system with a 532 nm incident laser. The Raman spectra were fitted using Lorenztian functions, providing the FWHM and area of the G and D peaks. The curve fitting results were then used to extract the density of line and point defects according to the theoretical simulations described in~\cite{canccado2017disentangling}. Details of the Raman analysis are given in Supporting Information. 

\noindent \textbf{FSCV measurements:} To examine the electrode sensitivity, FSCV measurements of dopamine and serotonin (Sigma Aldrich) were performed using a custom-made FSCV setup. The FSCV setup consists of a low-noise current amplifier (SR570, Stanford Research Systems), a data acquisition card (NI 6353 X series, National Instruments), and a MATLAB control interface. In FSCV sensing experiments, the biomolecules were dissolved in a 1x phosphate buffer saline (PBS) solution. PBS was prepared using the recipe in~\cite{chazotte2012labeling}. 

\bibliography{nanoengineered-graphene}
\bibliographystyle{ScienceAdvances}

\noindent \textbf{Acknowledgements:} 
This research used resources of the Center for Functional Nanomaterials, which is a U.S. DOE Office of Science Facility, at Brookhaven National Laboratory under Contract No. DE-SC0012704. We also acknowledge the Surface Science Facility of CUNY Advanced Science Research Center for the use of the XPS tool. DS acknowledges partial financial support by NSF-CMMI award 1728051. RK is supported by the National Institute of Mental Health grant R01MH109180, a Pew Scholarship in the Biomedical Sciences, and Simons Collaboration on the Global Brain.

\noindent \textbf{Author Contributions} T.W., A.A., D.S., R.K. designed research;  T.W., A.A., D.S. performed research; T.W., A.A., D.S., R.K. analyzed data; and T.W., A.A., D.S., R.K. wrote the paper.\\
\noindent \textbf{Competing Interests} The authors acknowledge the following patent applications: U.S. Serial No. 62/599,303 and U.S. Serial No. 62/539,045.\\

\clearpage

\clearpage

\clearpage

\clearpage

\clearpage

\end{document}